\documentclass[sigconf]{acmart}
\usepackage{xcolor}
\usepackage{cleveref}
\usepackage{listings}
\usepackage{array}
\lstdefinestyle{c}{
  frame=none,
  xleftmargin=2pt,
  stepnumber=1,
  numbers=left,
  numbersep=5pt,
  numberstyle=\ttfamily\tiny\color[gray]{0.3},
  belowcaptionskip=\bigskipamount,
  captionpos=b,
  escapeinside={*'}{'*},
  language=C,
  keepspaces=true,
  tabsize=2,
  keywordstyle=[1]{\color{purple}},
  keywordstyle=[4]{\color{blue}},
  keywordstyle=[5]{\color{purple!70!black}},
  morekeywords=[1]{
    auto, break, case, const, continue, default, do, else, enum,
    extern, for, goto, if, inline, register, restrict, return,
    signed, sizeof, static, struct, switch, typedef, union, unsigned,
    volatile, while, include, define, ifndef, endif
  },
  morekeywords=[4]{
    int, long, short, void, char, double, float, bool, size_t,
    uint8_t, uint16_t, uint32_t, uint64_t, int8_t, int16_t,
    int32_t, int64_t
  },
  commentstyle=\color{teal},
  stringstyle=\ttfamily,
  delim=[s][\ttfamily\color{orange}]{"}{"},
  columns=flexible,
  basicstyle=\small\ttfamily,
  showspaces=false,
  showstringspaces=false,
}
\lstdefinestyle{make}{
  frame=none,
  xleftmargin=2pt,
  stepnumber=1,
  numbers=left,
  numbersep=5pt,
  numberstyle=\ttfamily\tiny\color[gray]{0.3},
  belowcaptionskip=\bigskipamount,
  captionpos=b,
  language={[gnu]make},
  keepspaces=true,
  tabsize=2,
  keywordstyle=\color{purple},
  identifierstyle=\color{blue!70!black},
  commentstyle=\color{teal},
  stringstyle=\color{orange},
  columns=flexible,
  basicstyle=\small\ttfamily,
  showspaces=false,
  showstringspaces=false,
}
\lstdefinestyle{haskell}{
  frame=none,
  xleftmargin=2pt,
  stepnumber=1,
  numbers=left,
  numbersep=5pt,
  numberstyle=\ttfamily\tiny\color[gray]{0.3},
  belowcaptionskip=\bigskipamount,
  alsoletter={=<>-|{}},
  captionpos=b,
  escapeinside={*'}{'*},
  keywordstyle=[2]{\color{teal}},
  keywordstyle=[3]{\color{violet}},
  keywordstyle=[1]{\bfseries\color{purple}},
  keywordstyle=[4]{\color{blue}},
  keywordstyle=[5]{\color{cyan!50!black}},
  keywordstyle=[6]{\bfseries\color{purple}},
  keywords=[3]{
    INLINE, PRAGMA, NOINLINE, RULES, SPECIALISE, LANGUAGE, forall,
    DataKinds, TypeApplications
  },
  keywords=[4]{
    Int, Bool, Double, Maybe, Either, Monad, Map, String, ByteString,
    IO, Proxy, KnownSymbol, Show, Read, IORef, Num, MVar,
    Word32, Word64, Property, MonadIO, Type, FilePath, Vector,
    Integer, Setup, Secure, Nonsecure, NSEffects, SEffects, EXTI, GPIO,
    UART, Locked, Unlocked, Nil, Cons, N0, N1, N2, N3, N4, N7, N13,
    A, B, C, D, E, F, G, H, EXTIEdge, Member, Delete, Fresh, Periph,
    TZSC, RCC, GPIOActions, UARTActions, NonSecureCallable, Callable,
    NFData, True, False, Nothing, Just, Left, Right,
    UARTConfig, GPIOConfig, EXTIConfig,
    NONE, OUTPUT, INPUT, NOPULL, PULLUP, PULLDOWN, AF0,
    BOTH, FALLING, RISING, RCC_GPIOC, RCC_GPIOG, StateT, SetupState,
    Ptr, StablePtr, BFILE, ExtiCallback, CallableDummy, Word8, CInt, NonsecureEffects,
    COL0, COL1, COL2, ROW0, ROW1, ROW2, ROW3, KeyPos, GATE, LOCKOUT, SecureEffects,
    UnlockResult, Granted, Denied, LockedOut, Eq, UDB, Char
  },
  morekeywords=[1]{
    do, if, then, else, case, of, class, data, newtype, instance,
    where, deriving, import, let, in, module, qualified, type
  },
  morekeywords=[6]{div},
  otherkeywords={<.>, <-, ->, <>, ::, ':, \$, +, [, ],/,\\, \,, =, =>, ==>, |, \{, \}, (, ), -\#, \#-, @},
  morekeywords=[2]{-\#, \#-, \{, \#, \}},
  literate={<|>}{{\textcolor{purple}{\bfseries<|>}}}3,
  tabsize=2,
  keepspaces=true,
  comment=[l]{--},
  commentstyle=\color{teal},
  stringstyle=\ttfamily,
  showspaces=false,
  delim=[s][\ttfamily\color{orange}]{"}{"},
  columns=flexible,
  basicstyle=\small\ttfamily,
  showstringspaces=false,
}
\lstdefinestyle{haskellinline}{
  style=haskell,
  columns=fullflexible,
  basicstyle=\ttfamily,
  keywordstyle=[1]{\color{purple}},
}
\lstdefinestyle{cinline}{
  style=c,
  numbers=none,
  xleftmargin=0pt,
  basicstyle=\ttfamily,
  columns=fullflexible,
}
\NewDocumentCommand{\inlinec}{v}{\lstinline[style=cinline]|#1|}
\NewDocumentCommand{\inlinehaskell}{v}{\lstinline[style=haskellinline]!#1!}

\AtBeginDocument{%
  }

\begin{document}

\title{MicroHasTEE: Bare-Metal Haskell for Type-Level Peripheral Ownership on Armv8-M}
\subtitle{A Framework Realized on STM32 Nucleo Boards}

\author{Robert Krook}
\email{krookr@chalmers.se}
\affiliation{%
  \institution{Chalmers University of Technology and The University of Gothenburg}
  \city{Gothenburg}
  \country{Sweden}
}

\begin{abstract}
Arm TrustZone for Armv8-M isolates Secure and Non-secure software, but developers must still coordinate peripheral attribution, interrupt routing, initialization, and gateway interfaces across separately built firmware images.
Inconsistent assumptions between these images can compile successfully and emerge only as faults on the target device.
  
We present MicroHasTEE, a multiparty programming framework that expresses both firmware applications as participants in one typed Haskell program.
MicroHasTEE represents peripheral authority with type-level capability ledgers and uses indexed setup computations to track resource acquisition, configuration, transfer, and finalization.
Domain-specific effect types restrict peripheral operations and interrupt callbacks to the participant that holds the corresponding authority, while typed callable handles describe the Secure services available to Non-secure code.
MicroHs compiles the shared program twice to produce separate bare-metal Secure and Non-secure firmware images.
  
We implement MicroHasTEE for an STM32U5 Nucleo board, including TrustZone configuration, peripheral drivers, and a serialized gateway for cross-domain Haskell calls.
For programs expressed through its interface, MicroHasTEE rejects inconsistent resource use, attribution changes after configuration, callbacks in the wrong domain, and calls to unregistered Secure services.
A door-lock case study demonstrates feasibility, with firmware images occupying 232.7 KiB and 228.4 KiB of flash and approximately 220 KiB of SRAM per domain.
\end{abstract}




\maketitle

\section{Introduction}

Arm TrustZone for Armv8-M, or TrustZone-M, partitions a microcontroller into hardware-isolated Secure and Non-secure states and permits transitions through explicitly exposed Secure entry points~\cite{armTrustZoneM}.
TrustZone-M does not, however, determine which state owns each peripheral, GPIO pin, or SRAM region.
Vendors implement these assignments through device-specific security hardware.
On the STM32U5 Nucleo board used in this work, memory-mapped security registers, including those of ST's Global TrustZone Controller (GTZC), encode the security attribution of peripherals and SRAM.
Developers must therefore update this STM32-specific configuration when transferring a resource between states, keeping initialization, interrupt routing, and access permissions consistent with it.

Conventional STM32 development turns this configuration into a coordination problem between two separately built firmware images.
Each image can type-check while relying on assumptions about resource ownership that conflict with those of the other image.
Secure code may continue to use a UART after the firmware has assigned it to the Non-secure state, or Non-secure code may assume access to a GPIO pin that the Secure configuration never released.
The hardware enforces the configured boundary at run time, but it cannot determine whether the two applications agree about that boundary.
Developers therefore discover many configuration mismatches only when they execute the firmware on the target device.

Trusted Firmware-M (TF-M)\cite{tfm} is the reference implementation of the Platform Security Architecture's Secure Processing Environment for Armv8-M and Armv8.1-M systems.
It provides secure boot, isolated Secure partitions, standardized communication with Non-secure software, and common services such as cryptography, protected storage, and attestation.
TF-M Secure Partition manifests declare the services, memory-mapped peripherals, and interrupts required by each partition, while platform-specific code maps these declarations to the underlying security hardware.
TF-M therefore provides the principal conventional framework for constructing portable TrustZone-M systems.

MicroHasTEE addresses a complementary source-language problem.
Rather than describing Secure partitions and their resources in separate manifests and programs, it derives both firmware images from one typed program and checks peripheral operations against a shared account of resource authority and transfer.

MicroHasTEE treats TrustZone firmware as a multiparty Haskell program whose participants are the Secure and Non-secure applications.
Although these participants execute as separate firmware images, the programmer implements them simultaneously in one source-level program.
This program describes the computations performed by each participant, the resources available to them, the transfer of resources between them, and their permitted cross-boundary calls.
MicroHs~\cite{augustsson2024microhs} compiles the shared program separately for each TrustZone state.
Each resulting image contains the MicroHs runtime and executes compiled Haskell application code directly on the microcontroller without an operating system or RTOS.
A small C layer provides startup and raw peripheral access, while the application logic, resource partitioning, and cross-state interactions are expressed and executed as Haskell computations.
To the best of our knowledge, MicroHasTEE is the first framework to execute Haskell in both states of a bare-metal TrustZone-M microcontroller and to use a single typed Haskell program to coordinate their resource assignments and cross-state interactions.

This shared description lets MicroHasTEE check assumptions that would remain unrelated in separately written applications.
The type checker verifies that each participant uses only the resources currently available to it and that both participants agree on resource transfers.
It also checks cross-boundary calls against the Secure functions exposed to the Non-secure participant.
Both images are therefore checked against one account of their shared hardware and interface, rather than only being checked in isolation.
A disagreement becomes a compile-time error instead of a fault observed after deployment.

These guarantees apply to operations expressed through MicroHasTEE.
The framework cannot account for arbitrary C or FFI code that accesses a peripheral directly or modifies security registers outside its interface.
The MicroHasTEE programming model is not specific to the STM32U5.
It can target other TrustZone-M microcontrollers with sufficient flash and SRAM, provided that the platform establishes the required memory partition and that a MicroHasTEE backend implements the supported peripheral and interrupt attribution operations.
The implementation evaluated in this work uses the STM32U5-specific security hardware.

This work makes three contributions:

\begin{itemize}
  \item \textbf{A multiparty programming model for TrustZone-M firmware.}
  MicroHasTEE lets developers implement Secure and Non-secure applications together as participants in one source-level program and then compile that program into separate firmware images.

  \item \textbf{Static agreement about resources and cross-boundary interfaces.}
  MicroHasTEE checks each participant's resource use against a shared account of resource authority, transfers, and permitted calls, allowing it to reject inconsistencies before the firmware reaches the target device.

  \item \textbf{Bare-metal Haskell in both TrustZone-M states.}
  The prototype embeds the MicroHs runtime in each firmware image and executes compiled Haskell application code in both states without an operating system or RTOS.
  It connects the high-level shared resource description to the STM32 security configuration and hand-written peripheral drivers.
\end{itemize}

\section{Conventional Bare-Metal TrustZone-M Development}

To construct a conventional bare-metal TrustZone-M system, the programmer must express one security partition through several independent hardware and software mechanisms.
This section first introduces the mechanisms defined by Armv8-M and then explains how STM32 microcontrollers supplement them with device-specific controls for memory, peripherals, and interrupts.
Together, these mechanisms form the low-level programming model that MicroHasTEE seeks to make safer.

\subsection{Memory Partitioning}

\begin{figure*}[t]
  \centering
  \begin{minipage}[c]{0.43\textwidth}
    \centering
    \includegraphics[width=\linewidth]{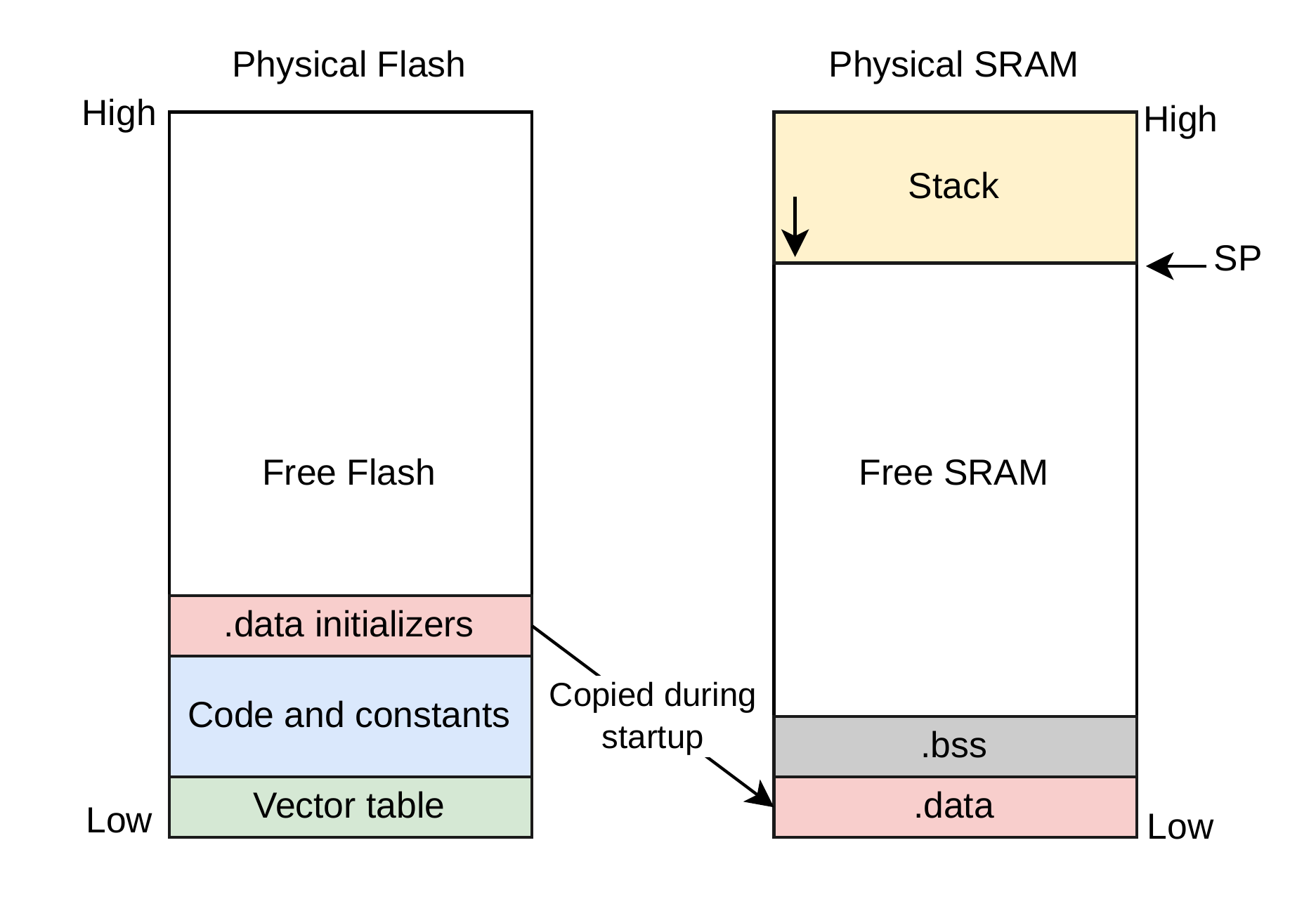}
    \par\smallskip
    \textbf{(a)} Single application
  \end{minipage}\hfill
  \begin{minipage}[c]{0.54\textwidth}
    \centering
    \includegraphics[width=\linewidth]{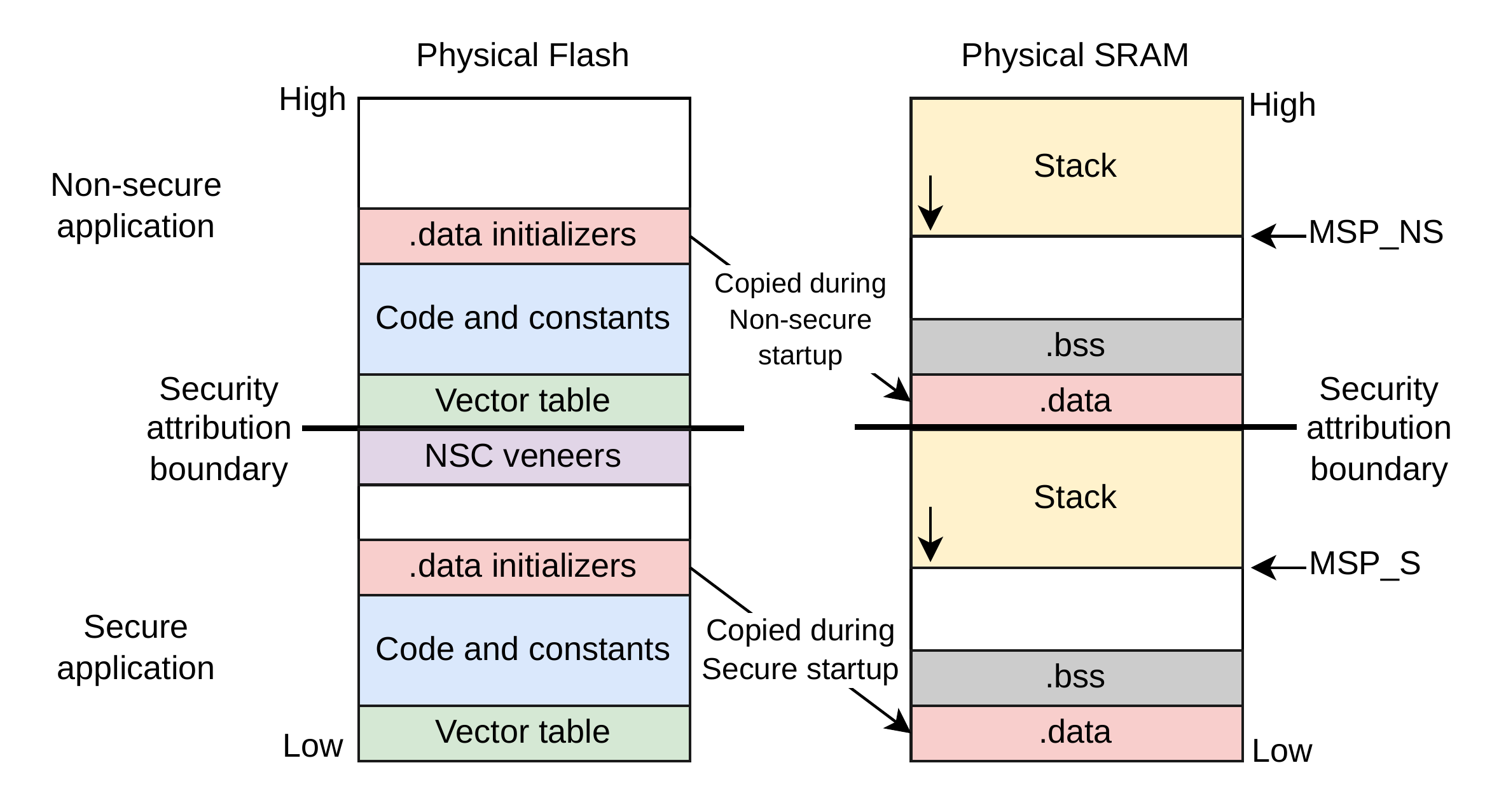}
    \par\smallskip
    \textbf{(b)} Secure and Non-secure applications
  \end{minipage}
  \caption{Physical flash and SRAM layouts for a single bare-metal application and two TrustZone-M applications.
  The layouts are not drawn to scale.}
  \Description{Side-by-side memory maps compare a single bare-metal application with Secure and Non-secure applications partitioned across the same physical flash and SRAM.}
  \label{fig:memory-partition}
\end{figure*}

A microcontroller contains fixed amounts of flash memory and SRAM, each of which may comprise several physical banks.
A linker script assigns a firmware image to address ranges and places its vector table, code, static data, and stack within them.
\Cref{fig:memory-partition}(a) shows this layout for a conventional bare-metal application.
Because Secure and Non-secure firmware are built as separate images, each image requires its own linker script.
The Secure linker script must also reserve a Non-secure Callable (NSC) region for the gateway veneers used to enter Secure code.

TrustZone-M allows developers to partition memory into Secure and Non-secure regions.
Our implementation assigns one primary flash range and one primary SRAM range to each security state and reserves a Non-secure Callable subregion within Secure flash.
The MCU's alignment and block-size requirements constrain the boundaries of these regions.
As \cref{fig:memory-partition}(b) shows, the two firmware images occupy disjoint regions of the same physical flash and SRAM.

The linker scripts describe this intended layout, but they do not enforce its security attributes.
Secure startup code must configure the architectural and STM32-specific attribution mechanisms to enforce the same partition.
The two images can therefore link successfully even when the linker and hardware configurations disagree.
Such a mismatch typically triggers a fault only when firmware accesses the affected region.

On the STM32U5, ST's \textit{Implementation Defined Attribution Unit} (IDAU) assigns fixed security attributes to address ranges~\cite{stm32u5ReferenceManual}.
The Cortex-M33 supplements this fixed mapping with a programmable \textit{Security Attribution Unit} (SAU).
For each CPU access, the processor combines the IDAU and SAU results and selects the more restrictive attribute.
The processor orders these attributes as Secure \(>\) Non-secure, Non-secure Callable \(>\) Non-secure~\cite{armTrustZoneM}.

The IDAU and SAU classify only transactions issued by the CPU, while other bus masters can access SRAM without passing through them.
ST therefore protects each SRAM bank at its memory interface with a block-based memory protection controller (MPCBB).
Each MPCBB assigns Secure or Non-secure attributes to individual SRAM blocks.
The SAU/IDAU and MPCBB configurations must assign compatible attributes to every SRAM region used by the firmware.
In our prototype, this memory partition is fixed by the linker scripts and Secure startup code.
MicroHasTEE does not represent flash or SRAM regions in its capability ledgers or reconfigure their attribution during application setup.
Instead, it assumes that this platform configuration correctly isolates the two firmware images.

\subsection{Peripheral Attribution with the GTZC}

Armv8-M defines the Secure and Non-secure states, memory attribution, and the mechanisms for transferring control between the two states.
STM32 microcontrollers supplement these architectural mechanisms with controls that assign security attributes to memory-mapped peripherals.
The \textit{Global TrustZone Controller} (GTZC) includes a \textit{TrustZone Security Controller} (TZSC), which configures the security attribution of supported peripherals~\cite{stm32u5ReferenceManual}.
Although the GTZC provides additional protection and monitoring mechanisms, we focus here on the TZSC's role in peripheral attribution.

Each peripheral controlled by the TZSC has a corresponding bit in a security configuration register.
A Secure attribution blocks Non-secure transactions, whereas a Non-secure attribution permits them.
A Non-secure attribution therefore does not give the Non-secure state exclusive access to the peripheral.

Some STM32 peripherals provide finer-grained attribution through their own security registers.
For example, each GPIO port controls up to 16 pins and provides a security configuration bit for each implemented pin.
Secure software can therefore retain Secure access to selected pins while permitting Non-secure access to other pins on the same port.
Only Secure software can modify these security attributes.

\subsection{Interrupt Attribution}

Peripheral attribution does not by itself determine where interrupts from that peripheral execute.
Armv8-M assigns each external interrupt a target security state through the Interrupt Target Non-secure State (\texttt{ITNS}) registers in the Nested Vectored Interrupt Controller (NVIC)~\cite{armTrustZoneM}.
Secure software configures this target, and the processor uses either the Secure or the Non-secure vector table when it takes the interrupt.

STM32 devices may also attribute the interrupt source independently of its NVIC target.
For example, to move a button input to the Non-secure application, Secure firmware must make the GPIO pin Non-secure; configure the corresponding EXTI line and mark it Non-secure; and target its NVIC interrupt to the Non-secure state.
The Non-secure image must also contain a handler in the corresponding slot of its vector table.
These settings must agree before the interrupt is enabled.
Releasing the GPIO pin alone is therefore insufficient: if the EXTI attribution, NVIC target, or vector entry disagrees, the event may be delivered to the wrong state or to a handler that cannot service it.

\subsection{Non-secure Callable Functions and Gateway Veneers}

The Secure and Non-secure firmware images occupy disjoint regions of flash and SRAM.
This isolation is asymmetric: Secure software can access Non-secure resources, whereas Non-secure software cannot directly access Secure memory or peripherals.
Secure firmware can nevertheless expose selected entry functions to Non-secure callers without exposing its remaining code and data.

Secure firmware declares an entry function using the CMSE \inlinec{cmse_nonsecure_entry} attribute~\cite{armCmse}.
The compiler marks the function as a Secure entry point, and the Secure linker creates a corresponding secure gateway veneer in a Non-secure Callable region.
The veneer consists of an \texttt{SG} instruction followed by a branch to the entry function in Secure memory.
Non-secure code calls the exported gateway symbol as an ordinary function.
When the processor executes \texttt{SG} at a valid gateway, it changes to the Secure state and continues to the entry function.
The compiler-generated epilogue clears register state as required and executes \texttt{BXNS} to return directly to the Non-secure caller.

Below is a simple function defined by the Secure application, with an attribute that marks it as an NSC function.

\begin{lstlisting}[style=c]
int __attribute__((cmse_nonsecure_entry))
add10(int value) {
  return value + 10;
}
\end{lstlisting}

The Non-secure application declares the function and calls it like an ordinary external function.
The linker will resolve it when the firmware is built.

\begin{lstlisting}[style=c]
extern int add10(int value);

int myadd(void) {
  return add10(5);
}
\end{lstlisting}

This declaration is sufficient to compile the Non-secure source, but it is not sufficient to link the Non-secure image.
The address associated with \inlinec{add10} is the address of the gateway veneer created when the Secure image is linked.
With the GNU toolchain, the Secure link uses \texttt{--cmse-implib} and \texttt{--out-implib} to emit a CMSE import library containing the exported symbols and their veneer addresses.
The Non-secure linker must then consume this generated library.

The simplified build dependency is therefore:

\begin{lstlisting}[style=make]
secure.elf:
	$(CC) ... -Wl,--cmse-implib,\
	            --out-implib=secure_cmse_import.lib

nonsecure.elf: secure.elf
	$(CC) ... secure_cmse_import.lib
\end{lstlisting}

The two images are linked separately, but they cannot be built independently.
The Secure link must run first, and any change to the exposed interface or veneer placement requires the import library and Non-secure image to be regenerated together.
Moreover, the C linker resolves \inlinec{add10} by symbol name and address; it does not verify that the independently compiled declarations on the two sides have matching types.

\begin{figure}[ht]
  \centering
  \includegraphics[width=\columnwidth]{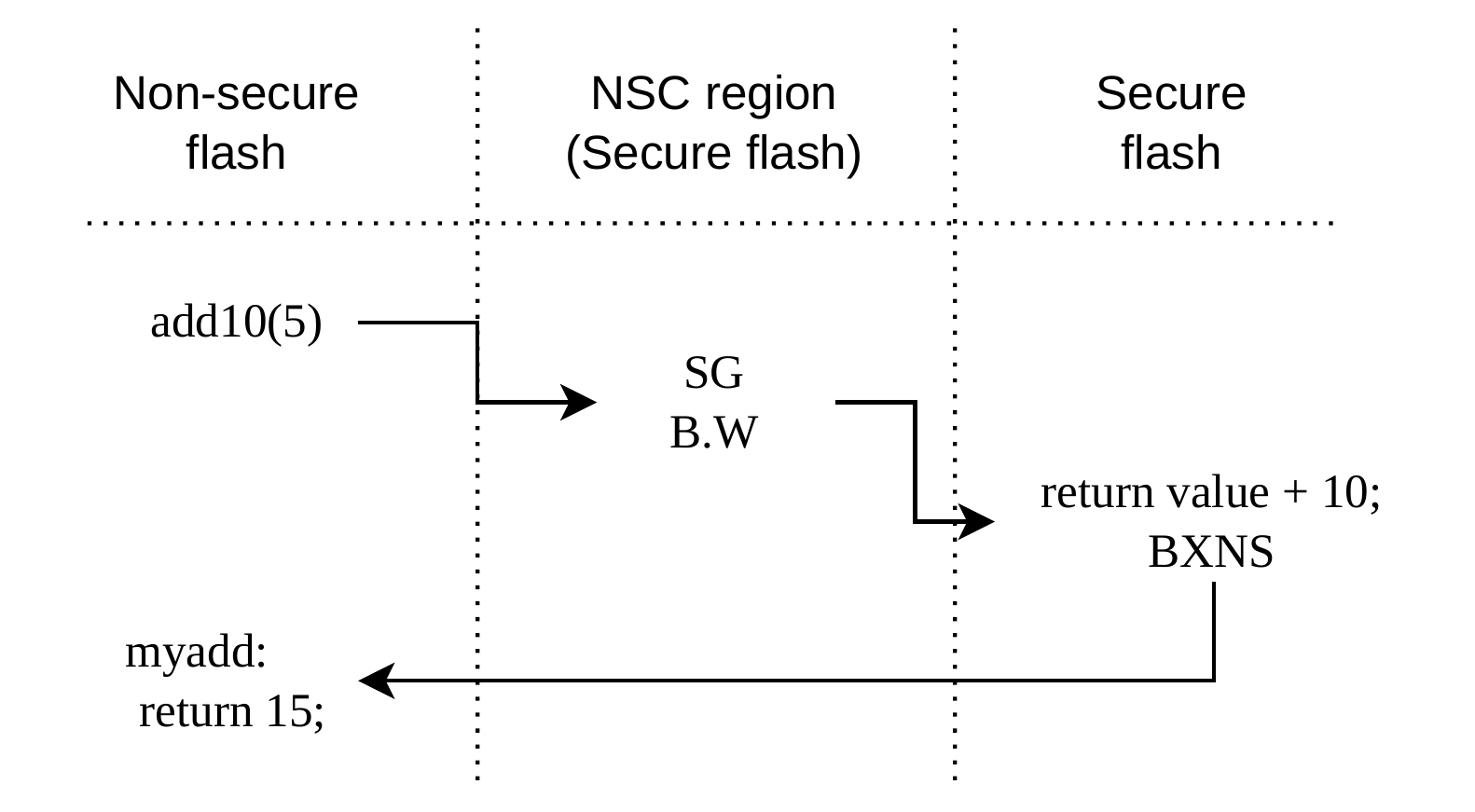}
  \caption{Control flow of a Non-secure call to a Secure entry function.
  The \texttt{SG} instruction in the NSC gateway veneer changes the processor to Secure state, after which \texttt{B.W} branches to the Secure function body.
  The function returns directly to its Non-secure caller with \texttt{BXNS}.}
  \Description{A three-column diagram shows a call from Non-secure flash entering an SG and B.W gateway veneer in an NSC region, executing the add10 function in Secure flash, and returning directly to the Non-secure caller with BXNS.}
  \label{fig:nsc-call}
\end{figure}

The control flow of this call from Non-secure to Secure code is shown in \cref{fig:nsc-call}.

A Secure entry function must treat pointer arguments received from Non-secure code as untrusted.
Before dereferencing a pointer, it must verify that the entire referenced range lies in Non-secure memory, has the permissions required by the operation, and does not wrap around the address space when its size is added~\cite{armCmse}.
After validating an input range, Secure code copies its contents into Secure memory and performs subsequent checks and computations on that copy, preventing later modifications to the original buffer from changing the value being processed.

This section describes several relatively small mechanisms.
The difficulty arises when these mechanisms must work together and agree on one security policy.
The simple act of allowing a Non-secure button handler to call a Secure service involves peripheral and interrupt attribution, a Non-secure vector entry, matching declarations in two programs, a gateway veneer, and an ordered build in which the Secure link produces input to the Non-secure link.
Each mechanism may by itself look correct, while some other part of the system disagrees with the policy.

\section{Threat Model and Scope}

MicroHasTEE addresses two distinct concerns.
Its type system detects configuration inconsistencies introduced by a trusted but fallible developer, while TrustZone-M provides run-time isolation from compromised Non-secure firmware.

We assume that matching firmware images are initially generated from the same MicroHasTEE program and deployed together.
The adversary may then execute arbitrary Non-secure code, read and modify Non-secure memory, use Non-secure-attributed peripherals, and invoke NSC gateways with arbitrary pointers and messages.
Any service registered as callable is available to the entire Non-secure domain, because callable handles are not authorization capabilities.
We assume that the adversary cannot modify Secure firmware, change security attribution, directly access Secure resources, or enter Secure code except through designated NSC gateways.
Our run-time objective is to preserve the confidentiality and integrity of Secure code, data, and peripherals, except for effects intentionally exposed by registered services.

MicroHasTEE uses a separate fault model for the developer.
For programs expressed through its public interface, the type system checks resource transfers, resource use, callback placement, and the registration of callable services.
After \inlinehaskell{lock_configuration}, well-typed code cannot perform further attribution-changing operations through that interface.
These guarantees do not apply to raw C, foreign-function calls, or internal modules, and the type-level ledgers express software authority rather than exclusive hardware access.

The guarantees assume that both builds execute corresponding setup operations in the same order and that MicroHasTEE, its hardware backend, the Secure gateway, the MicroHs toolchain and runtime, and the underlying hardware are correct.
Malformed gateway messages are not covered by the static types but remain within the run-time threat model.
The gateway must validate and copy Non-secure input before trusted deserialization code processes it.
We exclude physical attacks, fault injection, side channels, invasive debugging, firmware rollback, denial of service, and faults within the trusted computing base.

\section{MicroHasTEE}

MicroHasTEE treats the Secure and Non-secure applications as participants in one multiparty program.
This program describes the computations performed by each participant, the resources assigned to each participant, and the interactions allowed between them.
The build system compiles the program twice, once for each TrustZone domain.
The Secure build executes \inlinehaskell{Secure} computations and represents \inlinehaskell{Nonsecure} computations with placeholders.
The Non-secure build executes \inlinehaskell{Nonsecure} computations and represents \inlinehaskell{Secure} computations with placeholders.
Each build selects a different implementation of the same MicroHasTEE interface.
Both builds therefore type-check the same setup sequence while deriving the view required by their respective domains.
During setup, the program attributes tracked peripherals, GPIO pins, and interrupts, registers callbacks\footnote{Throughout this paper, \textit{callback} refers specifically to a handler triggered by an attributed GPIO pin's EXTI line, such as a button press. MicroHasTEE does not currently use the term for other interrupt sources, such as SysTick.} and Secure services, and starts the Non-secure application.
Before setup begins, the linker scripts and Secure startup code have already established the fixed flash and SRAM partition.

MicroHasTEE represents resource ownership with a pair of type-level capability ledgers, one for each participant.
The \inlinehaskell{Setup} type is an indexed monad~\cite{DBLP:journals/jfp/Atkey09} whose indices describe the ledgers before and after a setup computation.

\begin{lstlisting}[style=haskell]
  data Setup nsPre sPre nsPost sPost a
\end{lstlisting}

The parameters \inlinehaskell{nsPre} and \inlinehaskell{sPre} describe the Non-secure and Secure ledgers before the setup computation.
The parameters \inlinehaskell{nsPost} and \inlinehaskell{sPost} describe the corresponding ledgers afterward.
The final parameter, \inlinehaskell{a}, denotes the value returned by the computation.
Each public attribution operation updates the hardware configuration and performs the corresponding transition between ledger pairs.
Assuming a correct hardware backend and no access through raw C, FFI, or internal interfaces, the final pair therefore describes the attribution established during setup.
Subsequent MicroHasTEE operations require evidence that their resource occurs in the executing participant's ledger, so the compiler rejects operations for which that participant lacks authority.

The configuration phase always begins with the MicroHasTEE program being \textit{open to configuration}.\footnote{
For readability, the listings in this section abbreviate the type-level list representation with promoted-list syntax, such as \texttt{'[GPIO 7 C]}, rather than the \texttt{Cons (GPIO 7 C) Nil} representation used by the current MicroHs implementation.
They also write \texttt{Setup} computations with ordinary \texttt{do} notation, whereas the implementation uses \texttt{QualifiedDo} and \texttt{Ix.do} because \texttt{Setup} is an indexed monad.}
This is encoded by the capability ledger for the Secure domain carrying a \inlinehaskell{Unlocked} entry.
Consider the program below

\begin{lstlisting}[style=haskell]
  configure :: Setup '[] '[Unlocked] '[UART] '[Locked] ()
  configure = do
    uart <- get_console
    -- configuration of the actual UART device omitted
    tzsc_release_periph uart
    lock_configuration
\end{lstlisting}

The type of \inlinehaskell{configure} describes both the initial and final ledgers.
Initially, the Non-secure ledger is empty and the Secure ledger contains \inlinehaskell{Unlocked}.
The final ledgers assign the UART to the Non-secure participant and leave the Secure participant with no tracked peripheral capabilities.
The \inlinehaskell{Locked} entry records that the program may no longer change the configuration.
The following annotations show how each operation transforms the ledgers.

\begin{lstlisting}[style=haskell]
                 -- Initially ns = [],     s = [Unlocked]
  uart <- get_console      -- ns = [],     s = [UART, Unlocked]
  tzsc_release_periph uart -- ns = [UART], s = [Unlocked]
  lock_configuration       -- ns = [UART], s = [Locked]
\end{lstlisting}

The capability ledgers enforce that operations happen only when the ledger permits them.
For instance, MicroHasTEE requires that the UART is configured \textit{while it belongs to the Secure domain}.
Between lines 3 and 5 in the \inlinehaskell{configure} function, MicroHasTEE allows the UART to be configured.
The compiler rejects the program below, as the UART is configured \textit{after} it has been released to the Non-secure domain.

\begin{lstlisting}[style=haskell]
  configure :: Setup '[] '[Unlocked] '[UART] '[Locked] ()
  configure = do
    uart <- get_console

    tzsc_release_periph uart
    uart_init uart $ UARTConfig { baudrate    = 115200
                                , word_length = 8
                                , stop_bits   = 1
                                , parity      = NONE
                                }

    lock_configuration
\end{lstlisting}

\inlinehaskell{uart_init} requires both \inlinehaskell{Unlocked} and \inlinehaskell{UART} to be present in the Secure capability ledger.
At this point, the Secure ledger is \inlinehaskell{'[Unlocked]}.
Configuration remains open, but the UART is no longer available to Secure code, so the second constraint cannot be satisfied.

Acquiring a resource requires it to be absent from both ledgers.
After \inlinehaskell{tzsc_release_periph}, \inlinehaskell{UART} is absent from the Secure ledger but remains present in the Non-secure ledger.
The second \inlinehaskell{get_console} call is rejected by the compiler.

\begin{lstlisting}[style=haskell]
  configure :: Setup '[] '[Unlocked] '[UART] '[Unlocked] ()
  configure = do
    uart <- get_console
    tzsc_release_periph uart
    uart <- get_console
\end{lstlisting}

The restrictions demonstrated so far ensure that setup maintains a consistent account of what the security attribution on the hardware looks like.
The ledger stops representing the attribution of the program if the user can change it \textit{after} they have installed callbacks or invoked the application.
MicroHasTEE therefore provides a one-way transition that finalizes the capability ledger, disallowing any future modification.

\begin{lstlisting}[style=haskell]
  lock_configuration :: ( Member Unlocked s
                        , Delete Unlocked s s')
                     => Setup ns s ns (Locked ': s') ()
\end{lstlisting}

The \inlinehaskell{Member Unlocked s} constraint requires the Secure ledger to contain \inlinehaskell{Unlocked}.
The \inlinehaskell{Delete Unlocked s s'} constraint describes the ledger that remains after removing it.
The output ledger adds \inlinehaskell{Locked} to those remaining capabilities.
All operations that configure the device or change the ledgers require \inlinehaskell{Unlocked}.
Operations that install callbacks or register callable Secure services instead require \inlinehaskell{Locked}.
A program must therefore finish changing attribution before it performs operations whose types depend on the final ledgers.
Calling \inlinehaskell{lock_configuration} changes only the type-level setup state and performs no corresponding hardware operation.
The compiler rejects the following program because \inlinehaskell{get_gpio} attempts to acquire the red LED after configuration has been finalized.

\begin{lstlisting}[style=haskell]
  bad = do
    lock_configuration
    red <- get_gpio @2 @G -- rejected
\end{lstlisting}

Finalizing setup produces the ledgers against which the two participant computations are checked.
MicroHasTEE represents these computations with the \inlinehaskell{Secure} and \inlinehaskell{Nonsecure} monads.

\begin{lstlisting}[style=haskell]
  data Secure    effects a
  data Nonsecure effects a
\end{lstlisting}

Peripheral operations are overloaded across these monads.
For example, \inlinehaskell{gpio_toggle} is a method of the \inlinehaskell{GPIOActions} class.

\begin{lstlisting}[style=haskell]
  class GPIOActions m where
    gpio_toggle :: Member (GPIO pin port) effects
                => GPIO pin port -> m effects ()
\end{lstlisting}

MicroHasTEE provides \inlinehaskell{GPIOActions} instances for both \inlinehaskell{Secure} and \inlinehaskell{Nonsecure}.
In either instance, the \inlinehaskell{Member} constraint requires the GPIO pin to occur in the capability ledger attached to the selected monad.
The same operation can therefore be used by either participant, but only when that participant has authority over the pin.
Consider the two programs below

\begin{lstlisting}[style=haskell]
  goodNonsecure :: GPIO 7 C -> Nonsecure '[GPIO 7 C] ()
  goodNonsecure pin = gpio_toggle pin

  badSecure :: GPIO 7 C -> Secure '[Locked, GPIO 2 G] ()
  badSecure pin = gpio_toggle pin
\end{lstlisting}

The compiler accepts \inlinehaskell{goodNonsecure} because its Non-secure ledger contains \inlinehaskell{GPIO 7 C}.
It rejects \inlinehaskell{badSecure} because its Secure ledger contains only \inlinehaskell{Locked} and \inlinehaskell{GPIO 2 G}, not \inlinehaskell{GPIO 7 C}.

With these mechanisms explained, we next show how to install a callback.
If the user has properly configured and attributed a GPIO device and its corresponding EXTI device, a callback can be installed against it.
When the underlying GPIO pin receives an edge transition, the installed callback is invoked.
Consider the function for installing a Non-secure callback below.

\begin{lstlisting}[style=haskell]
  exti_on_nonsecure :: ( Member Locked s
                       , Member (EXTI pin port) ns)
                    => EXTI pin port
                    -> EXTIEdge
                    -> (EXTIEdge -> Nonsecure ns ())
                    -> Setup ns s ns s ()
\end{lstlisting}

The type requires attribution to have been finalised (as evident by the \inlinehaskell{Locked} being present in the Secure capability ledger), and requires the interrupt to belong to the Non-secure participant.
Furthermore, the callback itself is a computation in the \inlinehaskell{Nonsecure} monad, whose list of effects unifies against the Non-secure capability ledger of the \inlinehaskell{Setup} monad.
This is a good point to reiterate why we lock down the configuration.
If unification is successfully done here against the capability ledger of the Non-secure domain, we don't want the program to proceed and make further configurations.
This will alter the capability ledger, making the guarantees provided by the prior unification moot.

So far we have described how security attribution is configured, and how callbacks are installed.
Crucially, we have not yet shown how the Non-secure domain can invoke Secure services from the Secure domain.
This is done by explicitly registering \inlinehaskell{Secure} functions as \textit{callable}, and then providing a carefully typed interface for invoking them.
Consider the simple function below.

\begin{lstlisting}[style=haskell]
  add :: Int -> Int -> Secure effects Int
  add x y = return (x + y)
\end{lstlisting}

The function \inlinehaskell{add} takes two parameters and computes their sum in the \inlinehaskell{Secure} monad.
This function on its own cannot be invoked from the \inlinehaskell{Nonsecure} monad.
However, by marking it as \inlinehaskell{Callable}, we can.

\begin{lstlisting}[style=haskell]
  callable :: (Member Locked s, NonSecureCallable s f)
           => f -> Setup ns s ns s (Callable f)

  (<.>) :: NFData a => Callable (a -> b) -> a -> Callable b
\end{lstlisting}

The \inlinehaskell{Member Locked s} constraint permits a Secure service to be registered only after setup has been finalized.

The \inlinehaskell{NonSecureCallable s f} constraint accepts a Secure computation over the effect ledger \inlinehaskell{s}, or a function of arbitrary arity that eventually produces such a computation.
It therefore unifies the service's effect ledger with the final Secure ledger carried by the surrounding \inlinehaskell{Setup} computation.
For programs expressed through the public interface, a registered service can consequently use only the resources retained by the Secure participant.

The \inlinehaskell{callable} function returns an opaque \inlinehaskell{Callable} handle through which the Non-secure participant may refer to the registered service.
The \inlinehaskell{<.>} operator applies one argument to such a handle while preserving the remaining function type.

\begin{lstlisting}[style=haskell]
  f <- callable add
  let g = f <.> 10 <.> 5
\end{lstlisting}

The two handles have the following types.

\begin{lstlisting}[style=haskell]
  f :: Callable (Int -> Int -> Secure effects Int)
  g :: Callable (Secure effects Int)
\end{lstlisting}

The type of \inlinehaskell{g} shows that both arguments have been supplied and that the service is ready to be invoked with \inlinehaskell{sg}.

\begin{lstlisting}[style=haskell]
  sg :: Callable (Secure s a) -> Nonsecure ns a
\end{lstlisting}

Given a fully applied handle, \inlinehaskell{sg} invokes the corresponding Secure computation and returns its result in the \inlinehaskell{Nonsecure} monad.
The type of \inlinehaskell{sg} is straightforward.
Within MicroHasTEE's public Haskell interface, \inlinehaskell{sg} is the only operation that transfers explicit argument and result values between the participants.
When the call is made, the parameters are passed from the Non-secure domain to the Secure domain, and when the result is returned it is passed from the Secure domain to the Non-secure domain.

Finally, when the callbacks have been installed and the Secure services have been made callable, we invoke the Non-secure application.
This is done through the \inlinehaskell{nonsecure} function.

\begin{lstlisting}[style=haskell]
  nonsecure :: Member Locked s
            => Nonsecure ns () -> Setup ns s ns s ()
\end{lstlisting}

The type states that if the attribution phase has been locked down, we can run the Non-secure application.

A small example of a complete application is shown below.

\begin{lstlisting}[style=haskell]
  {-# LANGUAGE DataKinds #-}
  {-# LANGUAGE TypeApplications #-}

  module BlinkExample where

  import MicroHasTEE

  type NSEffects = '[GPIO 7 C]
  type SEffects = '[Locked, GPIO 2 G]

  toggleLED :: GPIO 2 G -> Secure SEffects ()
  toggleLED = gpio_toggle

  loop :: GPIO 7 C
       -> Callable (Secure SEffects ())
       -> Nonsecure NSEffects ()
  loop green toggle = do
    gpio_toggle green
    sg toggle
    systick_delay_ms 500
    loop green toggle

  app :: Setup '[] '[Unlocked] NSEffects SEffects ()
  app = do
    board_init
    board_configure_pll
    hz <- board_sysclk_hz
    systick_configure (hz `div` 1000)

    rcc <- get_rcc
    rcc_enable rcc RCC_GPIOG
    rcc_enable rcc RCC_GPIOC

    let output = GPIOConfig { mode = OUTPUT
                            , pull = NOPULL
                            , alternate = AF0
                            }

    red <- get_gpio @2 @G
    gpio_init red output

    green <- get_gpio @7 @C
    gpio_init green output
    gpio_release green

    lock_configuration
    toggle <- callable (toggleLED red)
    nonsecure (loop green toggle)

  main :: IO ()
  main = runSetup app
\end{lstlisting}

The Non-secure loop toggles the green LED directly and toggles the red LED by invoking the registered Secure function through \inlinehaskell{sg}.
The red LED belongs to the Secure domain, and can only be manipulated by the Secure application.

MicroHs generates one C source file from each of the two builds.
The build system compiles and links each generated file with the MicroHs runtime, startup code, C glue code, and the drivers for its domain to produce the corresponding firmware image.

For programs expressed through its public interface, MicroHasTEE prevents duplicate acquisition, use without ledger authority, configuration after finalization, callbacks attached to the wrong participant, and calls to Secure functions that were not explicitly registered.
These guarantees rely on the MicroHasTEE implementation, its hardware backend, and the deployment of matching firmware images.
Raw C, FFI, internal interfaces, and malformed gateway input remain outside the static model.

\section{Implementation}
\label{sec:implementation}

This section describes how MicroHasTEE turns a source program into the two firmware images deployed on a microcontroller.
Our prototype targets an STM32U5A5ZJQ microcontroller, which combines a Cortex-M33 processor implementing Armv8-M TrustZone with 4 MiB of flash and 2.5 MiB of SRAM.
The firmware runs bare metal, without an operating system or RTOS.
The prototype therefore includes the linker scripts, startup code, and board support needed to initialize the device and start the MicroHs runtimes.
Before either MicroHasTEE participant starts, Secure startup code configures the SAU and MPCBBs to realize the fixed memory layout described in \cref{fig:memory-partition}.
This configuration, together with the linker scripts and the IDAU mapping, forms part of the trusted platform backend rather than the typed MicroHasTEE setup.

\subsection{Program Partitioning and Compilation}

A MicroHasTEE source program contains computations for both the Secure and Non-secure participants, but the two participants execute in separate firmware images.
The build system therefore invokes MicroHs twice on the same source program.

In the Secure build, the preprocessor selects an implementation in which \inlinehaskell{Secure} computations contain executable \inlinehaskell{IO} actions and \inlinehaskell{Nonsecure} computations become placeholders.
The Non-secure build selects the complementary implementation: \inlinehaskell{Nonsecure} computations contain executable actions, while \inlinehaskell{Secure} computations become placeholders.
The public types remain the same in both builds, so each build checks the same setup sequence and the same division of resources and computations.

The two interpretations reside in separate implementation modules.
The \texttt{SECURE} build flag determines which module the common MicroHasTEE interface re-exports.

\begin{lstlisting}[style=haskell]
-- Effectful/Internal/Secure.hs
data Secure effects a = Secure (IO a)

data Nonsecure effects a = Nonsecure

-- Effectful/Internal/NonSecure.hs
data Secure effects a = Secure

data Nonsecure effects a = Nonsecure (IO a)
\end{lstlisting}

These definitions are alternatives and never coexist in one build.
Both modules expose the same abstract types and operations, while giving runtime behavior only to computations belonging to the target domain.

\begin{figure}[ht]
  \centering
  \includegraphics[scale=0.24]{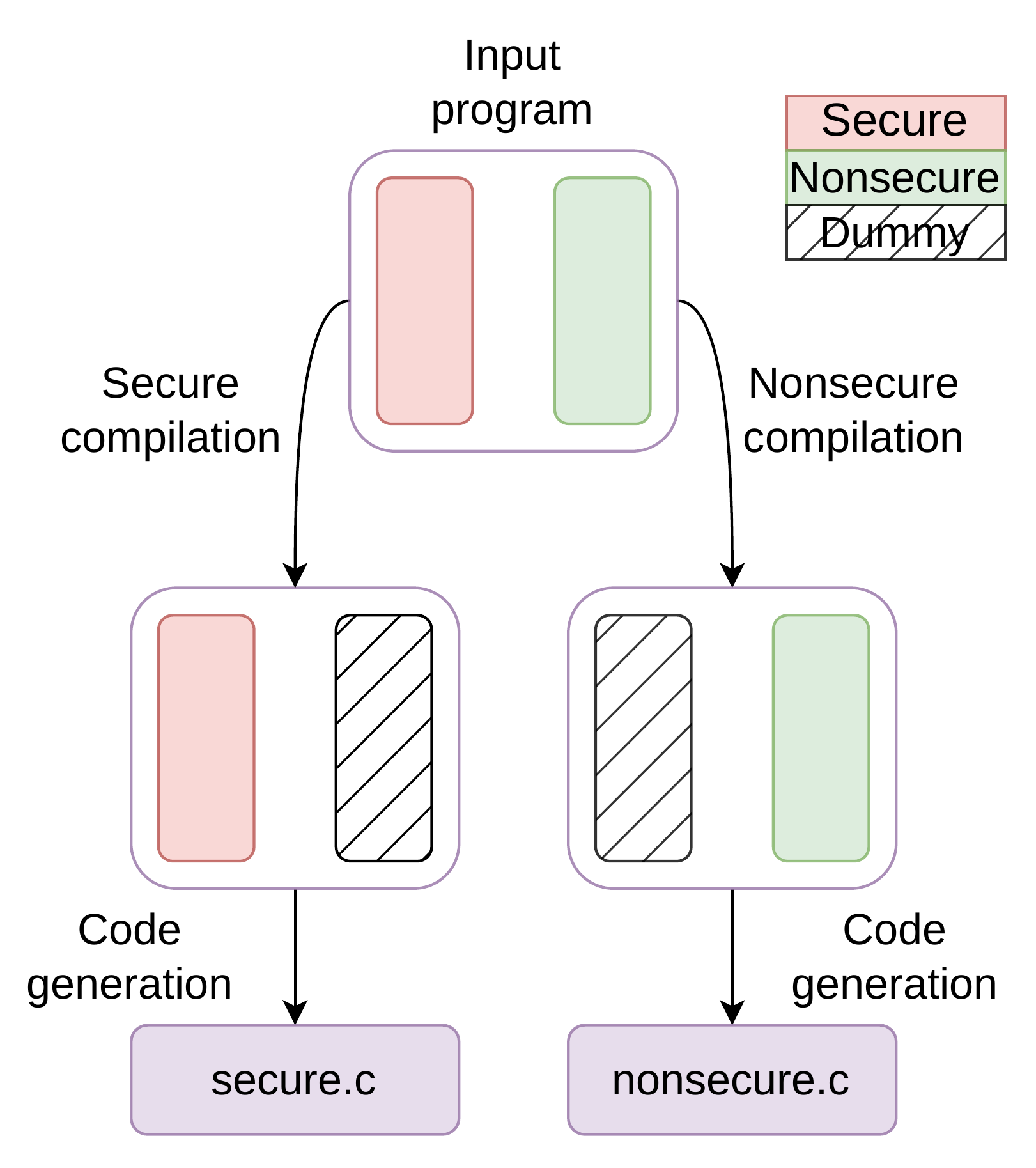}
  \caption{Compilation of one MicroHasTEE program into two domain-specific C programs.
The Secure build gives runtime meaning to Secure computations and replaces Non-secure computations with placeholders, while the Non-secure build does the converse.
MicroHs generates one C source file for each firmware image.}
  \Description{An input program containing Secure and Nonsecure regions is compiled twice. The Secure compilation keeps the Secure region and substitutes a dummy for the Nonsecure region; the Nonsecure compilation does the reverse. Code generation then produces secure.c and nonsecure.c respectively.}
  \label{fig:program-partition}
\end{figure}

\Cref{fig:program-partition} illustrates this process.
Each compilation retains the behavior associated with its target domain and erases the behavior of the other participant.
MicroHs then generates one C source file from each build.
The firmware build links each generated file with a separate MicroHs runtime, the startup code, the C gateway code, and the drivers compiled for that domain.
The result is a Secure firmware image and a Non-secure firmware image, flashed into their respective memory regions.

\subsection{Secure Function Registration and Invocation}

Although the participant computations have different runtime meanings in the two builds, both builds execute the same sequence of \inlinehaskell{Setup} operations.
The setup state contains the counter used to assign corresponding identifiers to callable functions.

\begin{lstlisting}[style=haskell]
  data Setup nsPre sPre nsPost sPost a =
    Setup (StateT SetupState IO a)
  
  data SetupState = SetupState
    { nonSecureCallable :: [( Int
                            , Ptr BFILE -> IO (Ptr BFILE)
                            )
                           ]
    , counter           :: Int
    , extiCallbacks     :: [ExtiCallback]
    }
\end{lstlisting}

Registration of a Secure callable function illustrates why the two executions of setup must remain synchronized.
Both builds assign each registration the current counter value and then increment the counter.
The Secure implementation uses \inlinehaskell{mkNSC} to convert the registered function into a dispatch handler, stores the identifier and handler in \inlinehaskell{nonSecureCallable}, and returns a placeholder handle.

\begin{lstlisting}[style=haskell]
  data Callable a = CallableDummy

  callable :: (Member Locked s, NonSecureCallable s a)
           => a -> Setup ns s ns s (Callable a)
  callable f = do
      let g inBf = mkNSC f inBf
      modify $ \st ->
          st { counter           = counter st + 1
             , nonSecureCallable =
                 (counter st, g) : nonSecureCallable st
             }
      return CallableDummy
\end{lstlisting}

The returned \inlinehaskell{Callable} value is a placeholder because the Secure firmware does not use the client-side handle.
In this build, the Non-secure computation that applies the handle and invokes \inlinehaskell{sg} is itself represented by a placeholder and is not executed.
On the Non-secure side, however, these operations carry a different meaning.

\begin{lstlisting}[style=haskell]
  data Callable a = Callable Int [Ptr BFILE -> IO ()]

  callable :: (Member Locked s, NonSecureCallable s a)
           => a -> Setup ns s ns s (Callable a)
  callable _ = do
      setupst <- get
      put $ setupst { counter = counter setupst + 1 }
      return $ Callable (counter setupst) []
\end{lstlisting}

In the Non-secure build, \inlinehaskell{callable} discards the Secure function because Secure computations have no runtime representation in that build.
It returns a handle containing the current numeric identifier and an initially empty list of argument serializers.
The correspondence between this identifier and the Secure dispatch table depends on both firmware images executing registrations in the same order.
MicroHasTEE therefore assumes that matching Secure and Non-secure images are built and deployed together.

After the Secure execution of setup completes, \inlinehaskell{runSetup} extracts the dispatch table from \inlinehaskell{SetupState}.
Gateway calls may occur after the setup computation has returned, so the implementation creates a \inlinehaskell{StablePtr} that keeps the table reachable by the Haskell runtime.
It stores the resulting opaque token in a Secure C global and retains it for the lifetime of the firmware.

Applying an argument to a callable function appends a writer action to the handle.
Given a pointer to a memory-backed \inlinehaskell{BFILE}, this action forces the argument to normal form and serializes it into the buffer.
Forcing the argument ensures that its evaluation occurs in the Non-secure state before serialization.

\begin{lstlisting}[style=haskell]
  (<.>) :: NFData a => Callable (a -> b) -> a -> Callable b

  Callable fun writes <.> x =
      Callable fun (writes ++ [\bf ->
        x `deepseq` primHSerialize bf x])
\end{lstlisting}

The MicroHs runtime provides serialization and deserialization primitives for supported Haskell values, so applications do not require hand-written serializers for each callable function.
MicroHasTEE uses these primitives to encode a call as a function identifier followed by its arguments.
Conceptually, the request and response that pass between the two domains have this format (with \texttt{||} denoting concatenation of byte strings).

\begin{lstlisting}[basicstyle=\small\ttfamily]
  request = serialize(identifier)
          || serialize(arg1)
          || ...
          || serialize(argN)

  response = serialize(result)
\end{lstlisting}

The MicroHs runtime format is a combinator graph, and serialized values can be quite large.
That is the chief reason for ensuring their evaluation before serialization, through the \inlinehaskell{NFData} type class.

The Non-secure implementation of the function \inlinehaskell{sg} produces this serialized package.
The code below shows the general operation of this function, but omits details concerning resource cleanup and error handling.
The internal function \inlinehaskell{nonsecureLiftIO} lifts the underlying \inlinehaskell{IO} action into the \inlinehaskell{Nonsecure} monad.
MicroHasTEE exposes neither this function nor a \inlinehaskell{MonadIO} instance to end users.

\begin{lstlisting}[style=haskell]
  sg :: Callable (Secure s a) -> Nonsecure ns a
  sg (Callable fun writeArgs) = nonsecureLiftIO $
    alloca $ \outLen ->
    allocaBytes resultBufferSize $ \outBuf -> do
      request <- c_openb_wr_mem -- fresh, Ptr BFILE
  
      -- create serialized package
      primHSerialize request fun
      mapM_ ($ request) writeArgs
  
      -- invoke the underlying NSC function
      c_sg request outBuf resultBufferSize outLen
  
      -- fetch the result
      len      <- peek outLen
      response <- c_openb_rd_mem outBuf len
      primHDeserialize response
\end{lstlisting}

The function \inlinehaskell{c_sg} is a Haskell FFI binding to the C function \inlinec{sg}, which the Secure image exposes through a gateway veneer in the Non-secure Callable region.
This is the prototype's only NSC entry point and the only direct cross-domain call path exposed by MicroHasTEE.

The Secure application performs the necessary validation checks, using \inlinec{cmse_check_address_range} to validate the request metadata and bytes as readable Non-secure memory and the result buffer and length field as writable Non-secure memory.
If all checks succeed, it immediately invokes the Secure Haskell runtime by calling a foreign exported Haskell function.
The example below shows this Haskell function, but omits details concerning error handling and resource management.

\begin{lstlisting}[style=haskell]
  handle_nsc_call :: Ptr Word8 -> CInt -> Ptr Word8
                  -> CInt -> Ptr CInt -> IO ()
  handle_nsc_call inBuf inLen outBuf outCapacity outLenPtr =
    do
      -- wrap the input bytes in a read-BFILE
      input   <- c_openb_rd_mem inBuf (fromIntegral inLen)

      -- deserialize the function index and look it up
      -- in the vtable
      funId <- primHDeserialize input :: IO Int

      table <- getVTable
      let fun = lookupFun funId table
      result <- fun input

      (resultBuf, resultLen) <- bfileBytes result
      if resultLen > fromIntegral outCapacity
        then poke outLenPtr (-1)
        else do
          copyBytes outBuf resultBuf resultLen
          poke outLenPtr (fromIntegral resultLen)
\end{lstlisting}

The \inlinehaskell{getVTable} operation reads the token from the Secure C global, reconstructs the \inlinehaskell{StablePtr}, and dereferences it to recover the dispatch table.
The receiver then uses the deserialized function identifier to select a handler from this table.

Every entry in the dispatch table has the uniform handler type
\inlinehaskell{Ptr BFILE -> IO (Ptr BFILE)}, although registered Secure functions may have different arities.
The \inlinehaskell{NonSecureCallable} instances construct this handler recursively from the function's type.
For a function of type \inlinehaskell{a -> b}, \inlinehaskell{mkNSC} deserializes one value of type \inlinehaskell{a}, applies the function to it, and continues with the resulting value of type \inlinehaskell{b}.
When the recursion reaches a \inlinehaskell{Secure effects r} computation, it executes the computation and serializes its result.
The type-class constraints determine the argument and result types during compilation, while the resulting dispatch table has a uniform runtime representation.

\subsection{Peripheral Drivers \& Building the Firmware}

\begin{figure}[ht]
  \centering
  \includegraphics[width=\columnwidth]{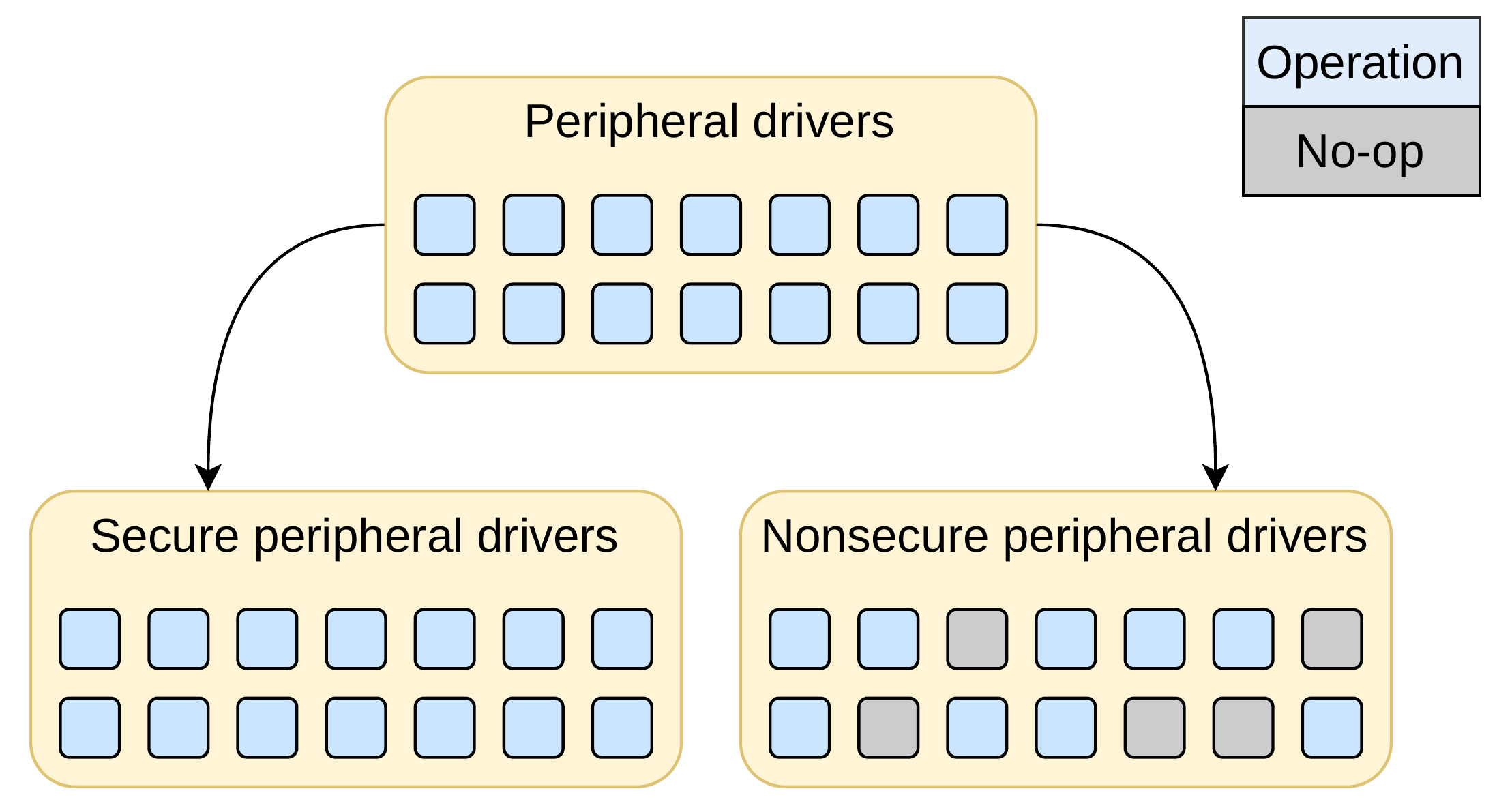}
  \caption{The peripheral driver library is split into its Secure and Non-secure builds; some operations become no-ops in the Non-secure build but remain exported to keep a unified API.}
  \Description{The peripheral driver library splits into a Secure and a Non-secure build. In the Secure build all operations are implemented; in the Non-secure build some operations become no-ops, shown as grey squares, while the rest remain implemented.}
  \label{fig:driver-partition}
\end{figure}

We have hand-written a library of peripheral drivers.
Like MicroHasTEE, the driver library is compiled twice, once for each domain, with identical APIs in both builds.
Some operations are guarded by CPP macros to only be included in the Secure build, such as operations for configuring security attribution.
The compilation mechanism for the peripheral drivers is illustrated in~\cref{fig:driver-partition}.

MicroHasTEE's typed HAL operations invoke this C driver library through the Haskell foreign-function interface.
Both builds type-check the same setup operations, but only the Secure driver implementation performs privileged attribution writes to mechanisms such as the TZSC, GPIO security registers, EXTI, and the NVIC.
The Non-secure archive retains the same function symbols while compiling privileged operations to no-ops.
The shared API therefore supports both interpretations of the setup program without allowing the Non-secure firmware to perform Secure configuration.

\begin{figure}[ht]
  \centering
  \includegraphics[width=\columnwidth]{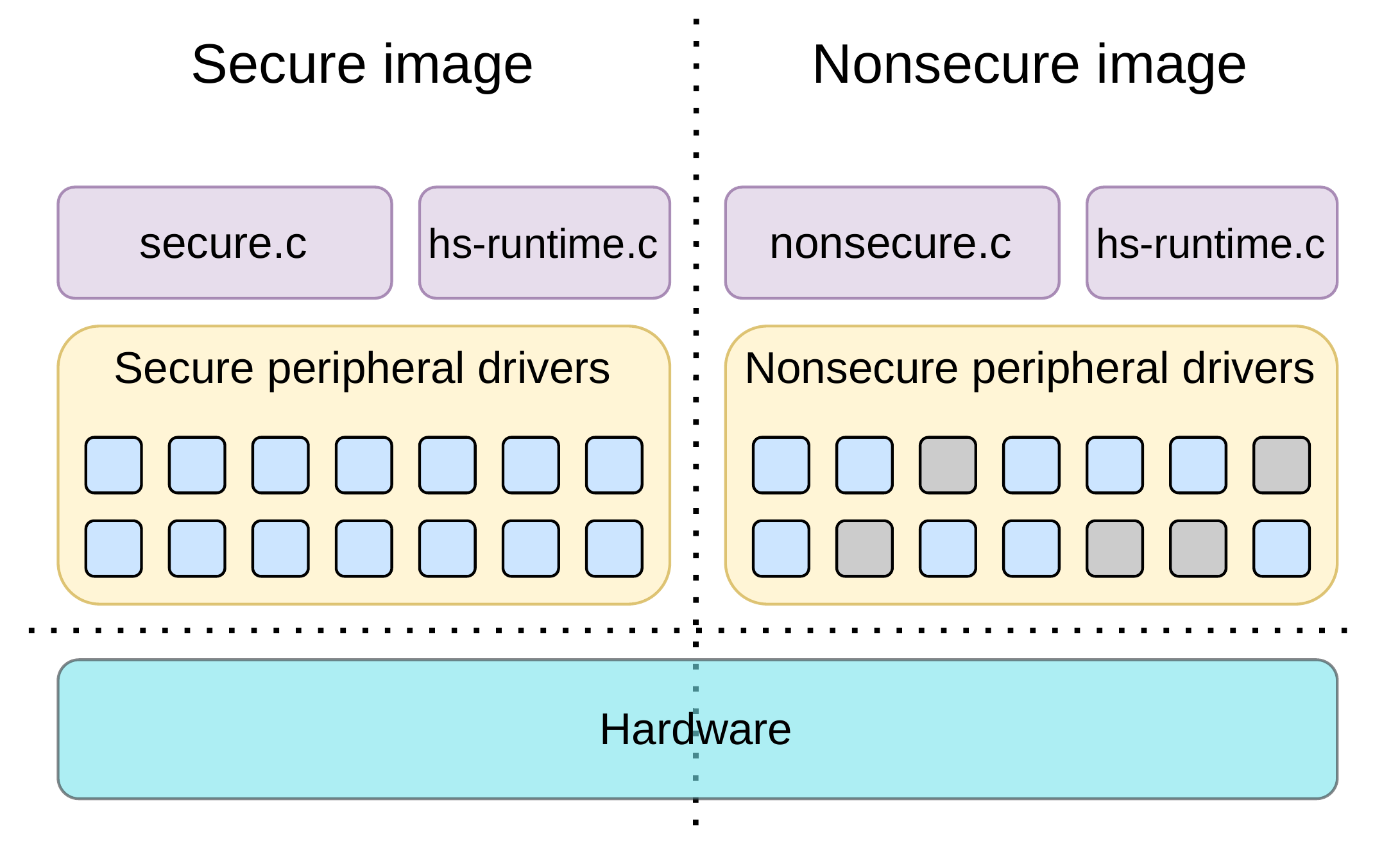}
  \caption{The Secure and Non-secure firmware images, each built from its generated MicroHasTEE C file and the MicroHs runtime on top of that domain's peripheral drivers, running on the shared hardware.
           The CMSE import library that the Secure build produces is omitted, as it is an artifact used by the linker, and is not present at runtime.}
  \Description{Two firmware images, Secure and Non-secure, each composed of a generated C file and the MicroHs runtime on top of that domain's peripheral drivers, both placed above a shared Hardware layer.}
  \label{fig:firmware-stack}
\end{figure}

Before the application can be flashed to the MCU, the complete firmware must be built.
First, the Secure image is built, producing a library file that the Non-secure image must link against to find the NSC entry points.
The Non-secure image is then built.
\Cref{fig:firmware-stack} shows an overview of the components that go into each of the firmware images, and how they are ordered.

\section{Evaluation}

We evaluate MicroHasTEE through a door-lock case study that exercises peripheral attribution, persistent Secure state, and a typed cross-boundary call.
We then identify the trusted computing base and measure the flash and SRAM requirements of the resulting firmware.

\subsection{Case Study: A Secure Door Lock}

The case study implements a PIN-controlled solenoid lock on an STM32U5A5ZJQ Nucleo board.
The user enters a four-character PIN on a \(4 \times 3\) matrix keypad.
A correct PIN energizes the solenoid for three seconds, while an incorrect PIN consumes one attempt.
After three consecutive failures, the Secure service rejects every subsequent request, including requests containing the correct PIN.

The Non-secure firmware owns the seven keypad GPIO pins and the UART used to report status.
The Secure firmware retains the solenoid-gate GPIO and the lockout-indicator GPIO.
It also stores the PIN and failed-attempt counter in a flash page reserved by the linker script.
Persistent storage prevents a power cycle from restoring an attacker's attempts.

The Secure solenoid-gate GPIO drives a MOSFET that switches the 6 V supply to the solenoid, while the seven Non-secure keypad GPIOs connect to the keypad's three columns and four rows.
We first describe the Secure service that enforces the access-control policy, then the Non-secure keypad client, and finally the shared setup that connects them.

\paragraph{Secure Application}

The Secure firmware implements the security-critical core of the application: it verifies submitted PINs, enforces the lockout policy, and controls the solenoid.
The Secure firmware owns two GPIO pins: \inlinehaskell{GATE} drives the MOSFET that controls the solenoid, while \inlinehaskell{LOCKOUT} drives the lockout indicator.

The \inlinehaskell{SecureEffects} list contains these GPIO capabilities and \inlinehaskell{Locked}, which records that security attribution has been finalized:

\begin{lstlisting}[style=haskell]
  type GATE    = GPIO 3 C
  type LOCKOUT = GPIO 2 G

  type SecureEffects = '[Locked, LOCKOUT, GATE]
\end{lstlisting}

Each unlock request returns an \inlinehaskell{UnlockResult} that distinguishes a successful unlock, an incorrect PIN, and a request rejected because the system is locked out:

\begin{lstlisting}[style=haskell]
  data UnlockResult
      = Granted
      | Denied Int   -- attempts remaining
      | LockedOut
      deriving (Show, Eq) -- NFData instance omitted
\end{lstlisting}

\inlinehaskell{Granted} indicates that the lock was actuated, \inlinehaskell{Denied n} reports the number of attempts remaining, and \inlinehaskell{LockedOut} indicates that the submitted PIN was not checked.

Three helpers implement the access-control policy.
Their types require the Secure GPIO capabilities used by each operation:

\begin{lstlisting}[style=haskell]
  grantAccess
      :: Member GATE effects
      => GATE -> Secure effects ()

  consumeAttempt
      :: Member LOCKOUT effects
      => UDB -> LOCKOUT -> Int
      -> Secure effects Int

  verifyAttempt
      :: (Member GATE effects, Member LOCKOUT effects)
      => UDB -> GATE -> LOCKOUT -> Int -> [Char]
      -> Secure effects UnlockResult
\end{lstlisting}

The function \inlinehaskell{grantAccess} pulses the solenoid gate.

The function \inlinehaskell{consumeAttempt} persists a failed attempt and activates the lockout indicator when the counter reaches \inlinehaskell{maxAttempts}.

The function \inlinehaskell{verifyAttempt} compares the submitted PIN with the stored PIN and selects between these operations.
The \inlinehaskell{Member} constraints require the corresponding GPIO capabilities to remain in the Secure ledger.

\begin{lstlisting}[style=haskell]
  door_unlock_attempt :: ( Member GATE effects
                         , Member LOCKOUT effects)
                       => UDB -> GATE -> LOCKOUT -> [Char]
                       -> Secure effects UnlockResult
  door_unlock_attempt db gate lockoutLed attempt = do
      count <- readLockoutCount db
      if count >= maxAttempts
          then do
              setLockoutIndicator lockoutLed True
              return LockedOut
          else verifyAttempt db gate lockoutLed count attempt
\end{lstlisting}

The \inlinehaskell{door_unlock_attempt} function first reads the persistent failure count.
If the count has reached \inlinehaskell{maxAttempts}, it activates the lockout indicator and returns \inlinehaskell{LockedOut} without checking the submitted PIN; otherwise, it delegates to \inlinehaskell{verifyAttempt}.
All of these functions execute in the \inlinehaskell{Secure} monad, but that fact alone does not expose them to the Non-secure firmware.
During shared setup, \inlinehaskell{callable} registers \inlinehaskell{door_unlock_attempt} as a Secure service and returns a \inlinehaskell{Callable} handle that the Non-secure firmware invokes through \inlinehaskell{sg}.

\paragraph{Non-secure Application}

The Non-secure firmware polls the keypad, collects four key presses, and uses a \inlinehaskell{Callable} handle to submit the resulting PIN to the Secure service.
The following type aliases identify the GPIO pins connected to the keypad.

The \inlinehaskell{NonsecureEffects} list records that these pins and the UART are available to Non-secure computations:

\begin{lstlisting}[style=haskell]
  type ROW0 = GPIO 11 E
  type ROW1 = GPIO 8  G
  type ROW2 = GPIO 7  G
  type ROW3 = GPIO 13 E
  type COL0 = GPIO 14 F
  type COL1 = GPIO 9  E
  type COL2 = GPIO 15 F

  type NonsecureEffects = '[ COL2, COL1, COL0
                           , ROW3, ROW2, ROW1
                           , ROW0, UART]
\end{lstlisting}

A \(4 \times 3\) matrix keypad uses seven GPIO pins rather than twelve.
Reading one column input does not identify a key by itself, because the key's position depends on both its row and its column.
The Non-secure firmware therefore drives the rows in turn while sampling the columns.
The function \inlinehaskell{scanRow} implements this behaviour.

\begin{lstlisting}[style=haskell]
  scanRow :: Member (GPIO pin port) NonsecureEffects
          => GPIO pin port -> COL0 -> COL1 -> COL2
          -> Nonsecure NonsecureEffects (Maybe Int)
  scanRow rowGpio col0 col1 col2 = do
      gpio_write rowGpio False
      c0 <- gpio_read col0
      c1 <- gpio_read col1
      c2 <- gpio_read col2
      gpio_write rowGpio True
      return $ if not c0 then Just 0
               else if not c1 then Just 1
               else if not c2 then Just 2
               else Nothing
\end{lstlisting}

The \inlinehaskell{scanRow} function returns the zero-based index of the first active column, or \inlinehaskell{Nothing} if that row contains no pressed key.
The \inlinehaskell{scanMatrix} function applies \inlinehaskell{scanRow} to each row until it detects a key, then combines the row and column indices into a \inlinehaskell{KeyPos}.
The rest of the Non-secure application uses \inlinehaskell{scanMatrix} to record key presses.
When it has collected four keys, it supplies the resulting code to the callable Secure service.
The following helper isolates the cross-boundary call:

\begin{lstlisting}[style=haskell]
  submitAttempt
      :: Callable
           ([Char] -> Secure SecureEffects UnlockResult)
      -> [Char]
      -> Nonsecure NonsecureEffects UnlockResult
  submitAttempt unlockFn code =
      sg (unlockFn <.> code)
\end{lstlisting}

After setup supplies the database and Secure GPIO handles, \inlinehaskell{unlockFn} accepts only the PIN argument.
Applying \inlinehaskell{code} with \inlinehaskell{<.>} supplies the final argument, making the handle ready for \inlinehaskell{sg}.
The function \inlinehaskell{sg} invokes the registered service and returns its result to the Non-secure caller.

\paragraph{Shared Setup}

The shared setup initializes the peripherals, releases the UART and keypad pins to the Non-secure ledger, and retains the gate and lockout-indicator pins in the Secure ledger.
After initializing the persistent database and Non-secure keypad state, it connects the two applications:

\begin{lstlisting}[style=haskell]
  lock_configuration

  unlockFn <- callable $
      door_unlock_attempt db gate lockoutLed

  irq_enable

  nonsecure $
      runKeypad uart stateRefAction unlockFn
                row0 row1 row2 row3
                col0 col1 col2
\end{lstlisting}

The call to \inlinehaskell{lock_configuration} fixes both effect ledgers before \inlinehaskell{callable} registers the Secure service.
The setup partially applies \inlinehaskell{door_unlock_attempt} to the database and Secure GPIO handles, leaving the submitted PIN as its only argument.
Finally, \inlinehaskell{nonsecure} passes the resulting handle and the released peripheral handles to \inlinehaskell{runKeypad}.
Together, the effect ledgers and callable interface let the Non-secure firmware read the keypad and request an unlock decision without receiving either Secure GPIO handle.

\subsection{Trusted Computing Base}

The Trusted Computing Base (TCB) is the set of software that must be correct for MicroHasTEE's security guarantees to hold.
A smaller TCB is desirable, since more code typically means more bugs.
The TCB of MicroHasTEE depends on not just the MicroHasTEE-specific software, but also on the arm-none-eabi-gcc compiler, libc, linker, and assembler.
However, for the sake of presentation we omit them and report only the components MicroHasTEE adds to the TCB.
The contents of the TCB are listed in~\cref{table:tcb}.

\begin{table}[ht]
  \caption{Source-code size of the components in MicroHasTEE's trusted computing base.}
  \label{table:tcb}
  \centering
  \begin{tabular}{lr}
    \toprule
    Component & SLoC \\
    \midrule
    Peripheral driver library & 1588 \\
    Startup code & 360 \\
    Linker scripts & 192 \\
    \addlinespace
    MicroHs compiler & 10137 \\
    MicroHs runtime system & 8644 \\
    MicroHs base-library source closure (conservative) & 13258 \\
    \addlinespace
    MicroHasTEE Haskell Code & 2068 \\
    MicroHasTEE C code & 340 \\
    \midrule
    Total & 36587 \\
    \bottomrule
  \end{tabular}
\end{table}

Of the peripheral drivers, just under 500 SLoC are header definitions whereas the remaining ~1k SLoC are actual source code.
The MicroHs base-library source closure is large, at just over 13k SLoC.
This is a conservative estimate, however, and the final dependency is most likely much lower.
This is the sum of lines of code in the transitive closure of all base-library modules that are loaded during compilation of MicroHasTEE.
However, even if we end up using just one or two definitions from a module, its entire content is counted in the reported number.
The compiler mitigates this by pruning unused definitions before generating code.

We observe that the code produced \textit{for} MicroHasTEE consists of \(4548\) SLoC, while the MicroHs compiler and runtime it relies on contribute the rest.
Out of the \(2068\) SLoC of Haskell code, roughly \(1500\) SLoC are bindings for the underlying peripheral drivers and the effect-tracking wrapper around them.
The actual MicroHasTEE multiparty programming framework is surprisingly small, consisting of ~400 SLoC.

\subsection{Memory Requirements}

The MicroHs runtime uses graph reduction to execute a program.
When a program is executed, the runtime system first constructs a graph in memory from a serialized representation emitted by the MicroHs compiler.
This graph contains both the program's code and its data, and is modified repeatedly during execution by a byte code interpreter.
The graph is updated in place to enable sharing.

\begin{table}[ht]
  \caption{Flash footprint of the Secure and Non-secure images of the MicroHasTEE door-lock example (STM32U5), attributed per component via the linker map.
           The C library row is code pulled in transitively from the toolchain (\texttt{libc}/\texttt{libgcc}) rather than written for this project.}
  \label{table:flash-requirement}
  \centering
  \begin{tabular}{lcr}
    \toprule
    Component & Secure (KiB) & Non-secure (KiB) \\
    \midrule
    MicroHs runtime & 75.9 & 75.9 \\
    C library (\texttt{libc}/\texttt{libgcc}) & 72.3 & 72.0 \\
    Example program & 67.9 & 67.0 \\
    Peripheral drivers & 7.6 & 5.1 \\
    udb (flash KV store) & 5.0 & 5.0 \\
    Bootloader / TZ init & 2.6 & 2.2 \\
    Board glue & 1.5 & 1.3 \\
    \midrule
    Total & 232.7 & 228.4 \\
    \bottomrule
  \end{tabular}
\end{table}

When the runtime is compiled for the MCU in this paper, it deliberately omits certain functionality.
As an example, the runtime's support for accepting the initial serialized program in a compressed format is not included, as it is not a feature we need.
The runtime system itself requires ~76 KiB of flash memory, and is roughly the same on both images, totaling 152 KiB.

Additionally, the serialized version of the initial program exists in a \texttt{const} array and is similarly compiled and stored in flash.
The example program used in the case study requires 67.9 KiB in the Secure image, and 67 KiB in the Non-secure image.

The peripheral drivers occupy much less, but also show more of a difference.
This is due to how they are designed, where a noticeable number of peripheral operations (configuration and security attribution) are only carried out by the Secure image.
As a result, the Non-secure peripheral library occupies 2.5 KiB less flash.

SRAM requirements are more difficult to assess, because memory usage is hard to predict in a lazy functional language.

The graph that MicroHs uses is built up of uniformly sized \textit{nodes}.
On a 32-bit build that we use for MCUs, a node occupies \(8\) bytes.
The runtime heap-allocates cells in which nodes live, as well as a stack of node pointers.
Each stack entry occupies 4 bytes.

While it is somewhat application specific how many heap cells a program will require, for the door-lock case study used in this paper, we allocate \(24000\) heap cells and a stack with room for \(5000\) node pointers.
We allocate \(750\) words for the garbage collector (~3 KiB), and require an additional ~10 KiB for other static globals declared in the \texttt{.data} and \texttt{.bss} sections.

In total, one firmware image will reliably require \textit{at least} ~220 KiB of SRAM at runtime.
If the program runs out of free heap cells, the MicroHs runtime does not try to allocate more and simply crashes.
The Secure and Non-secure images both require this amount of memory, so across both domains our application requires ~440 KiB of SRAM.
The STM32U5A5ZJQ we use has 2.5 MiB of SRAM, well above our requirements.

\section{Related Work}

\paragraph{TF-M}
TrustZone-M by itself gives nothing more than two address spaces and a fault when execution in one space reaches into the other.
TF-M~\cite{tfm} uses these mechanisms to implement a full trusted execution environment.
MicroHasTEE does not claim to be a trusted execution environment.
Instead, it adopts the minimal model TrustZone-M provides on its own.
MicroHasTEE models the two domains and ensures they remain separated, except through designated NSC gateways.
MicroHasTEE checks this separation at compile time, and checks that a program's assumed peripheral attribution matches its actual configuration, rejecting ill-formed programs before the firmware reaches the target device.
\Cref{table:tfm-vs-microhastee} collects some of the differences between TF-M and MicroHasTEE.

\begin{table}[ht]
  \caption{Comparison of TF-M and MicroHasTEE along several architectural and security properties.
\textit{Isolated Secure Partitions} refers to TF-M's ability to host multiple mutually isolated partitions within the Secure state; MicroHasTEE currently supports only a single Secure application.}
  \label{table:tfm-vs-microhastee}
  \centering
  \small
  \begin{tabular}{@{}p{0.42\linewidth} >{\centering\arraybackslash}p{0.24\linewidth} >{\centering\arraybackslash}p{0.24\linewidth}@{}}
    \toprule
    Property & TF-M & MicroHasTEE \\
    \midrule
    Secure boot and update & Yes & No \\
    Standardized security services (crypto, protected storage, attestation, update) & Yes & No \\
    Isolated Secure Partitions & Yes & No \\
    Resource manifests & Yes & No \\
    Joint Secure/Non-secure source & No & Yes \\
    Cross-domain resource agreement checked & Runtime fault & Compile-time \\
    Cross-boundary call checked against callee & Runtime (handle+iovec) & Compile-time (typed) \\
    Application language & C & Haskell (type- and memory-safe) \\
    Maturity & Production ready & Research prototype \\
    \bottomrule
  \end{tabular}
\end{table}

Two rows in \Cref{table:tfm-vs-microhastee} deserve further comment.

TF-M's PSA Client API identifies a callee and a request through an integer, similar to how MicroHasTEE registers a callable service under a numeric identifier that a caller later supplies.
However, \inlinec{psa_call} passes its arguments as an array of iovecs whose count and sizes are runtime values, so TF-M has no static means of checking that a caller supplies the number and size of arguments a service expects.
MicroHasTEE instead makes the argument list part of a registered service's type, so the compiler rejects a call with the wrong arity or argument types before the firmware is built.

TF-M partitions declare the peripherals, memory-mapped regions, and interrupts they require through a manifest, and the toolchain uses this manifest alone to derive the platform's secure attribution configuration, including the GTZC where present.
This manifest describes only the Secure partition; it is oblivious to the Non-secure application, and nothing ties its claims to what the Non-secure build assumes.
A developer can therefore still write Non-secure code that assumes access to a peripheral, such as a UART, that a partition's manifest has already claimed as Secure, and the mismatch surfaces only when the firmware executes.
MicroHasTEE closes this gap by checking both participants against one shared, compile-time account of attribution, turning such a disagreement into a compile-time error rather than a fault discovered on hardware.

\paragraph{Rust}
Rust provides experimental TrustZone-M support for Armv8-M targets through an unstable CMSE calling convention~\cite{rustCmse}.
For a function declared with the \texttt{cmse-nonsecure-entry} ABI, \texttt{rustc} emits the required entry symbol and Secure return sequence, and the linker uses that symbol to generate a Secure gateway veneer.
This support covers the compiler and linker boundary but does not configure security attribution.
Developers must still configure the architectural SAU and device-specific mechanisms such as the STM32 GTZC, peripheral attribution, and interrupt attribution~\cite{armTrustZoneM,stm32u5ReferenceManual}.
As with C, they must also coordinate the Secure and Non-secure firmware images and their shared interface.

\paragraph{HasTEE}
HasTEE\cite{DBLP:conf/haskell/SarkarKRC23} is a Haskell framework for confidential computing on Intel SGX.
Its multiparty programming model inspired MicroHasTEE's own design.
HasTEE targeted SGX enclaves on conventional machines, such as laptops, whereas MicroHasTEE targets TrustZone-M on a microcontroller.
HasTEE also relied on a patched GHC runtime system to run Haskell inside the enclave, whereas MicroHasTEE compiles and runs the upstream MicroHs runtime system on bare metal in both TrustZone-M states.

HasTEE's partitioning model builds on Haste.App, a multitier Haskell framework that assigns computations to client and server tiers~\cite{DBLP:conf/haskell/EkbladC14}.
MicroHasTEE adopts the same broad structure, with hardware-defined Secure and Non-secure tiers.
Choreographic programming instead treats endpoints symmetrically and describes their communication explicitly in a global program.
HasChor embeds this approach in Haskell and obtains per-endpoint behaviour through endpoint projection~\cite{DBLP:journals/pacmpl/ShenKK23}.
Its original implementation does not statically remove all code belonging to other endpoints; Krook and Hammersberg show how GHC rewrite rules and specialisation can partition both HasChor and Haste.App programs~\cite{DBLP:conf/haskell/KrookH24}.
MicroHasTEE differs from both approaches by making hardware-resource attribution and gateway authority, rather than general network communication, its central statically tracked properties.\\

There are relatively few projects where a high-level managed language like Haskell is used to run code in TrustZone-M.
If we instead look at TrustZone-A or Intel SGX, we find more such projects.

\paragraph{GoTEE}
GoTEE\cite{DBLP:conf/usenix/GhosnLB19} extends the Go compiler and runtime with support for \textit{secure routines}, in contrast to Go's ordinary goroutines.
GoTEE creates a secure routine inside an Intel SGX enclave.
GoTEE also implements channels that let the enclave communicate with the rest of the program running outside it.
The extended compiler partitions a program automatically, but does not help the programmer keep sensitive data from leaving the enclave.

\paragraph{J\(_{E}\)}
J\(_{E}\)\cite{DBLP:conf/csfw/OakABS21} implements a subset of Java and supports running programs inside enclaves.
It uses information-flow control to keep data from flowing through the program in undesirable ways.
A program first undergoes partitioning, and a second compiler pass then checks that the interactions between the two resulting programs are not ill-formed.

\paragraph{LispBM}
LispBM~\cite{lispbm} is a noteworthy example of a functional language, Lisp,  whose interpreter is written specifically for MCUs, without necessarily targeting a trusted execution environment.
It optimizes for performance in a resource-constrained environment and implements many features and extensions not commonly found in Lisp.
The VESC community has widely adopted it, using it to program motor controllers and build UIs for physical displays.

\paragraph{Synchron}
Synchron provides a functional-language virtual machine and high-level API for concurrent embedded programs on STM32F4 and NRF52 microcontrollers~\cite{DBLP:conf/ecoop/SarkarSS22}.
MicroHasTEE likewise executes a managed functional language on a microcontroller, but it targets paired TrustZone-M images and statically tracks their division of peripheral authority.

\section{Conclusions \& Future Work}

For programs expressed through its public interface, MicroHasTEE checks that peripheral attribution, resource use, interrupt placement, and cross-domain interfaces agree before the firmware is deployed.
The door-lock case study illustrates this by assigning capabilities for the keypad and UART to the Non-secure participant, while the GPIO controlling the solenoid is attributed Secure.
The type checker rejects programs whose configuration is inconsistent, catching bugs before deployment.

By analysing the case study we show that the programming framework is feasible on the STM32U5A5ZJQ used in this work.
Both firmware images require roughly 230 KiB of flash each, and 220 KiB of SRAM.
These requirements fit comfortably within the 4 MiB of flash and 2.5 MiB of SRAM of the device used in this work, but still indicate that a somewhat well-provisioned device is necessary.

Our evaluation covers one application on one STM32U5 device and exercises only a subset of the implemented peripherals.
It provides neither a formal soundness proof nor measurements of execution time, energy consumption, or real-time behaviour.
The results support claims about static consistency, feasibility, and memory requirements, but do not establish production readiness.

Further evaluation should characterize gateway latency, worst-case execution time, and energy consumption against an equivalent C implementation.
MicroHasTEE nevertheless demonstrates that a shared typed program can make cross-domain peripheral agreement a compile-time property on bare-metal TrustZone-M.

\begin{acks}
  We thank Bo Joel Svensson for patiently answering our many questions about embedded systems, and Lennart Augustsson for many helpful discussions about MicroHs and for his generous assistance with the compiler and runtime.
\end{acks}

\bibliographystyle{ACM-Reference-Format}
\bibliography{main}

@String{BIT = "{BIT}" }

@String{Computer = "{IEEE} Computer" }

@inproceedings{DBLP:conf/haskell/SarkarKRC23,
  author       = {Abhiroop Sarkar and
                  Robert Krook and
                  Alejandro Russo and
                  Koen Claessen},
  editor       = {Trevor L. McDonell and
                  Niki Vazou},
  title        = {HasTEE: Programming Trusted Execution Environments with Haskell},
  booktitle    = {Proceedings of the 16th {ACM} {SIGPLAN} International Haskell Symposium,
                  Haskell 2023, Seattle, WA, USA, September 8-9, 2023},
  pages        = {72--88},
  publisher    = {{ACM}},
  year         = {2023},
  url          = {https://doi.org/10.1145/3609026.3609731},
  doi          = {10.1145/3609026.3609731},
  bibsource    = {dblp computer science bibliography, https://dblp.org}
}

@inproceedings{augustsson2024microhs,
  author       = {Lennart Augustsson},
  editor       = {Niki Vazou and
                  J. Garrett Morris},
  title        = {MicroHs: {A} Small Compiler for Haskell},
  booktitle    = {Proceedings of the 17th {ACM} {SIGPLAN} International Haskell Symposium,
                  Haskell 2024, Milan, Italy, September 6-7, 2024},
  pages        = {120--124},
  publisher    = {{ACM}},
  year         = {2024},
  url          = {https://doi.org/10.1145/3677999.3678280},
  doi          = {10.1145/3677999.3678280},
  bibsource    = {dblp computer science bibliography, https://dblp.org}
}

@software{tfm,
  title        = {Trusted Firmware-M ({TF-M})},
  author       = {{Trusted Firmware Project}},
  organization = {Linaro},
  note         = {Retrieved August 2026},
  url          = {https://www.trustedfirmware.org/projects/tf-m/},
  urldate      = {2026-08-16}
}

@inproceedings{DBLP:conf/usenix/GhosnLB19,
  author       = {Adrien Ghosn and
                  James R. Larus and
                  Edouard Bugnion},
  editor       = {Dahlia Malkhi and
                  Dan Tsafrir},
  title        = {Secured Routines: Language-based Construction of Trusted Execution
                  Environments},
  booktitle    = {Proceedings of the 2019 {USENIX} Annual Technical Conference, {USENIX}
                  {ATC} 2019, Renton, WA, USA, July 10-12, 2019},
  pages        = {571--586},
  publisher    = {{USENIX} Association},
  year         = {2019},
  url          = {https://www.usenix.org/conference/atc19/presentation/ghosn},
  bibsource    = {dblp computer science bibliography, https://dblp.org}
}

@inproceedings{DBLP:conf/csfw/OakABS21,
  author       = {Aditya Oak and
                  Amir M. Ahmadian and
                  Musard Balliu and
                  Guido Salvaneschi},
  title        = {Language Support for Secure Software Development with Enclaves},
  booktitle    = {34th {IEEE} Computer Security Foundations Symposium, {CSF} 2021, Dubrovnik,
                  Croatia, June 21-25, 2021},
  pages        = {1--16},
  publisher    = {{IEEE}},
  year         = {2021},
  url          = {https://doi.org/10.1109/CSF51468.2021.00037},
  doi          = {10.1109/CSF51468.2021.00037},
  bibsource    = {dblp computer science bibliography, https://dblp.org}
}

@article{DBLP:journals/jfp/Atkey09,
  author       = {Robert Atkey},
  title        = {Parameterised notions of computation},
  journal      = {J. Funct. Program.},
  volume       = {19},
  number       = {3-4},
  pages        = {335--376},
  year         = {2009},
  url          = {https://doi.org/10.1017/S095679680900728X},
  doi          = {10.1017/S095679680900728X},
  bibsource    = {dblp computer science bibliography, https://dblp.org}
}

@inproceedings{DBLP:conf/haskell/EkbladC14,
  author       = {Anton Ekblad and
                  Koen Claessen},
  editor       = {Wouter Swierstra},
  title        = {A seamless, client-centric programming model for type safe web applications},
  booktitle    = {Proceedings of the 2014 {ACM} {SIGPLAN} symposium on Haskell, Gothenburg,
                  Sweden, September 4-5, 2014},
  pages        = {79--89},
  publisher    = {{ACM}},
  year         = {2014},
  url          = {https://doi.org/10.1145/2633357.2633367},
  doi          = {10.1145/2633357.2633367},
  bibsource    = {dblp computer science bibliography, https://dblp.org}
}

@article{DBLP:journals/pacmpl/ShenKK23,
  author       = {Gan Shen and
                  Shun Kashiwa and
                  Lindsey Kuper},
  title        = {HasChor: Functional Choreographic Programming for All (Functional
                  Pearl)},
  journal      = {Proc. {ACM} Program. Lang.},
  volume       = {7},
  number       = {{ICFP}},
  pages        = {541--565},
  year         = {2023},
  url          = {https://doi.org/10.1145/3607849},
  doi          = {10.1145/3607849},
  bibsource    = {dblp computer science bibliography, https://dblp.org}
}

@inproceedings{DBLP:conf/haskell/KrookH24,
  author       = {Robert Krook and
                  Samuel Hammersberg},
  editor       = {Niki Vazou and
                  J. Garrett Morris},
  title        = {Welcome to the Parti(tioning) (Functional Pearl): Using Rewrite Rules
                  and Specialisation to Partition Haskell Programs},
  booktitle    = {Proceedings of the 17th {ACM} {SIGPLAN} International Haskell Symposium,
                  Haskell 2024, Milan, Italy, September 6-7, 2024},
  pages        = {27--40},
  publisher    = {{ACM}},
  year         = {2024},
  url          = {https://doi.org/10.1145/3677999.3678276},
  doi          = {10.1145/3677999.3678276},
  bibsource    = {dblp computer science bibliography, https://dblp.org}
}

@inproceedings{DBLP:conf/ecoop/SarkarSS22,
  author       = {Abhiroop Sarkar and
                  Bo Joel Svensson and
                  Mary Sheeran},
  editor       = {Karim Ali and
                  Jan Vitek},
  title        = {Synchron - An {API} and Runtime for Embedded Systems},
  booktitle    = {36th European Conference on Object-Oriented Programming, {ECOOP} 2022,
                  Berlin, Germany, June 6-10, 2022},
  series       = {LIPIcs},
  volume       = {222},
  pages        = {17:1--17:29},
  publisher    = {Schloss Dagstuhl - Leibniz-Zentrum f{\"{u}}r Informatik},
  year         = {2022},
  url          = {https://doi.org/10.4230/LIPIcs.ECOOP.2022.17},
  doi          = {10.4230/LIPICS.ECOOP.2022.17},
  bibsource    = {dblp computer science bibliography, https://dblp.org}
}

@manual{armTrustZoneM,
  author       = {{Arm Limited}},
  title        = {{Arm TrustZone Technology for the Armv8-M Architecture}},
  number       = {{ARM 100690\_0201\_00\_en}},
  edition      = {Version 2.1},
  year         = {2018},
  month        = oct,
  url          = {https://developer.arm.com/documentation/100690/0201},
  urldate      = {2026-09-02}
}

@manual{stm32u5ReferenceManual,
  author       = {{STMicroelectronics}},
  title        = {{STM32U5 Series Arm-Based 32-Bit MCUs: Reference Manual}},
  number       = {{RM0456}},
  edition      = {Revision 6},
  year         = {2025},
  month        = jan,
  url          = {https://www.st.com/resource/en/reference_manual/rm0456-stm32u5-series-armbased-32bit-mcus-stmicroelectronics.pdf},
  urldate      = {2026-09-02}
}

@manual{armCmse,
  author       = {{Arm Limited}},
  title        = {{Armv8-M Security Extensions: Requirements on Development Tools}},
  edition      = {Version 1.4},
  year         = {2024},
  month        = jun,
  note         = {Issue date: 21 June 2024},
  url          = {https://arm-software.github.io/acle/cmse/cmse.html},
  urldate      = {2026-09-02}
}

@software{lispbm,
  author       = {Joel Svensson and LispBM contributors},
  title        = {{LispBM}},
  year         = {2026},
  note         = {Software repository, accessed September 2, 2026},
  url          = {https://github.com/svenssonjoel/lispBM},
  urldate      = {2026-09-02}
}

@misc{rustCmse,
  author       = {{The Rust Project Developers}},
  title        = {{The Rust Unstable Book: cmse\_nonsecure\_entry}},
  year         = {2026},
  howpublished = {\url{https://doc.rust-lang.org/unstable-book/language-features/cmse-nonsecure-entry.html}},
  note         = {Accessed September 2, 2026},
  urldate      = {2026-09-02}
}

\appendix

\end{document}